# EDITORIAL

## Memristor – Why Do We Have to Know About It?

When William Shockley proposed the bipolar transistor theory, nobody knew how to build such a device and the announcement about the transistor's invention in late 1948 by Bell labs did not evoke any public or professional interest. But now we know that the invention of bipolar transistor led to the beginning of the integrated circuit revolution. Today's integrated circuits contain more than a billion transistors of nanoscale dimensions, packed into a small area (about 120 mm²) and they have changed the way we live and communicate with each other. Integrated circuits contain both active devices, such as transistors, and passive devices - resistors (*R*), capacitors (*C*) and even inductors (*L*). We are so familiar with these three passive circuit elements that none of us ever thought that there could be a fourth circuit element other than *R*, *L* and *C*.

In 1971, a little known professor, Lean Chua, working at University of California, Berkeley, published a paper in the *IEEE Transactions on Circuits Theory*, postulating the existence of a fourth circuit element, in addition to those which we all know [1]. Chua named this element memristor. He even suggested that we may have to rewrite the circuit theory books, due to the existence of this fourth circuit element. No one actually paid any serious attention to this basic scientific discovery because there was no physical example or model for a memristor. However, in an article in *Nature*, when a group of scientists from Hewlett-Packard announced in May 2008 that they had built a prototype of a memristor and that its operation could be understood using a simple physical model, it led to a frenzied interest in Chua's work and his memristor [2]. The objective of this editorial is to provide a brief outline of this fourth circuit element, its operation and possible applications, so that we can appreciate why there is so much of interest in this scientific discovery.

All of us who have studied either physics or electrical engineering in college are aware that there are four circuit parameters: charge *q*, current *i*, voltage *v* and magnetic flux *φ*. The physical law that relates charge and current is

$$\frac{dq}{dt} = i \tag{1}$$

Similarly, the physical law relating flux and voltage is

$$\frac{d\varphi}{dt} = v \tag{2}$$

The above two relations are depicted in Figure 1. Besides, as shown in Figure 1, voltage and current are related by the following equation:

$$\frac{dv}{di} = R \tag{3}$$

Charge and voltage are related by:

$$\frac{dq}{dv} = C \tag{4}$$

Flux and current are related by:

$$\frac{d\varphi}{di} = L \tag{5}$$

But what about the relationship between flux *φ* and charge *q*? Can they also be related? This was the question examined by Chua in his 1971 paper. His elaborate circuit analysis, which is not easy to understand unless you have studied circuit theory deeply, describes the relationship between flux and charge by a simple equation -

$$\frac{d\varphi}{dq} = M \tag{6}$$

where *M* is defined as memristance, the property of a memristor just as the resistance is the property of a resistor. With this new relationship suggested by Chua, we will have six equations relating the four fundamental circuit parameters – *R*, *L*, *C* and the new found *M*. It will be very easy to visualize the inevitable presence of the memristor, if we rewrite eqs (3) – (6) as shown in the Table 1 [3]:

We see from the Table 1 that the integral can be used in four different ways to describe the relationship between current and voltage by either using it or not using it. We note that the equations for resistance and memristance appear identical, except for the presence of the integral sign in the latter's case on both sides of '='. However, this integral cannot be cancelled because the constant of integration need not be zero [3]. And this is the constant that makes the memristor 'remember' the previous state. The integral of current over time is nothing but charge and the integral of voltage over time is the magnetic flux φ or flux linkage. Therefore, we see that there are three siblings of Ohm's law and not just two, as we have been thinking all along.

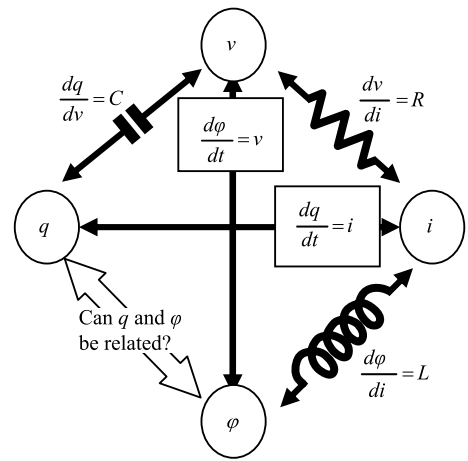

**Figure 1:** Possible relations among charge *q*, current *i*, voltage *v* and magnetic flux φ.

We know that the circuit components R, L and C are linear elements, unlike a diode or a transistor, which exhibit a nonlinear current-voltage behaviour. However, Chua has proved theoretically that a memristor is a nonlinear element because its current-voltage characteristic is similar to that of a Lissajous pattern. If a signal with certain frequency is applied to the horizontal plates of an oscilloscope and another signal with a different frequency is applied to the vertical plates, the resulting pattern we see is called the Lissajous pattern. A memristor exhibits a similar current-voltage characteristic, as shown in Figure 2. Unfortunately, no combination of nonlinear resistors, capacitors and inductors can reproduce this Lissajous behaviour of the memristor. That is why a memristor is a fundamental element.

How do we understand the meaning of memristance? According to Stanley Williams, the co-developer of memristor prototype, 'Memristance is a property of an electronic component. If charge flows in one direction through a circuit, the resistance of that component of the circuit will increase, and if charge flows in the opposite direction in the circuit, the resistance will decrease. If the flow of charge is stopped by turning off the applied voltage, the component will 'remember' the last resistance that it had, and when the flow of charge starts again the resistance of the circuit will be what it was when it was last active'[4]. In other words, a memristor is 'a device which bookkeeps the charge passing its own port' [5]. This ability to remember the previous state made Chua call this new fundamental element a memristor – short form for memory and resistor.

Are there any examples of a memristor in our everyday lives?

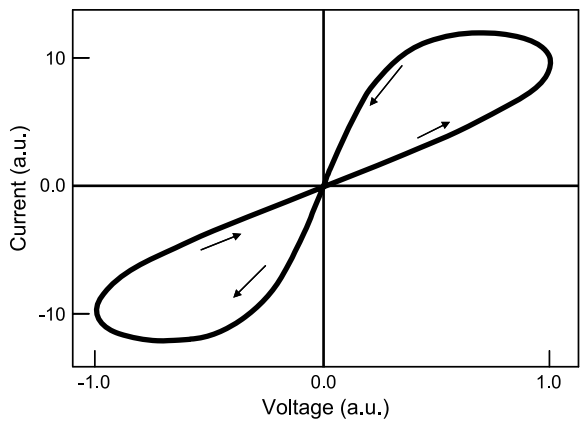

**Figure 2:** Typical current-voltage characteristics of a memristor.

Table 1: Various possible relationships between I and V

| Name | Law | Constant k | Its name |
|---|---|---|---|
| Resistor | I = kV | k = 1/R | Resistance |
| Capacitor | ∫I = kV | k = C | Capacitance |
| Inductor | I = k∫V | k = 1/L | Inductance |
| Memristor | ∫I = k∫V | k = 1/M | Memristance |

We have all seen the analog of a memristor at some time in our profession. The simplest of one such example is the electrolytic cell [5]. Any electrolytic cell has an internal resistance. However, when a current passes through it, the chemical reactions inside the electrolytic cell will affect the electrolyte concentration, making the resistance of the cell a function of the current passing through the electrodes of the cell. When the current is removed, the resistance of the cell will remain 'frozen'. In other words, the resistance of the cell is a function of the charge passing between the electrodes. With a similar charge controlled behaviour as that of an electrolytic cell, the memristor is a nonlinear element and can be described by the following mathematical relation [2]:

$$v = R[q(t)]\, i \qquad (7)$$

The meaning of this equation is that the charge flowing through the memristor dynamically changes the internal state of the memristor making it a nonlinear element.

What implications does the discovery of memristor hold for the future of electronics? If it is such an important circuit element, why has it taken nearly four decades to construct a prototype of the memristor? In history, we have several such instances of practical realizations lagging behind the theoretical conceptualization - the most famous being the case of the field effect transistor. The field effect concept was known as early as 1930, even before the invention of the bipolar transistor. However, the field effect transistor became a practical reality only in the early 60's, though it has now become an indispensable part of CMOS (complementary metal-oxide-semiconductor) technology. In the early days, we simply could not make a field effect transistor due to an imperfect interface between Si and $SiO_2$, which resulted in huge unacceptable levels of surface states blocking the field from controlling the carrier flow in the channel of a MOSFET (metal-oxide-semiconductor field-effect-transistor). In the case of memristor too, although the concept looked interesting, we did not know how to physically realize a memristor. Incidentally, there are dozens of scientific publications in the literature involving thinfilm devices, which exhibited a memristor-like current-voltage characteristics, but those scientists simply could not explain this 'strange behaviour'. Perhaps, they were unaware of Chua's discovery of the memristor. To see the first prototype realization of a memristor and understand its behaviour using a physical model, we had to wait for nanotechnology to evolve and for a team of scientists from Hewlett-Packard to realize that the device they had built was nothing but the missing memristor. Certain effects become 'visible' only at nanoscale! This is true in the case of memristor because the ideal model suggested by Chua indicates that the nonlinear behaviour of memristor increases by several orders of magnitude when the device dimensions are reduced to the nanoscale. That brings us to the question: what exactly did the Hewlett-Packard team do and how did they explain the memristor behaviour? The entire saga is very well explained in a cover story of *IEEE Spectrum* [6] by Stanley Williams. I strongly suggest that you read the article. However, let me explain their work very briefly.

Semiconductors are doped to make them either p-type or n-type. For example, if silicon is doped with arsenic, it becomes n-type. However, when we apply an electric field to a piece of n-type silicon, the ionized arsenic atoms sitting inside the silicon lattice will not move. We do not want them to move, in any case. Pure titanium dioxide ($TiO_2$), which is also a semiconductor, has high resistance, just as in the case of intrinsic silicon, and it can also be doped to make it conducting. If an oxygen atom, which is negatively charged, is removed from its substitutional site in $TiO_2$, a positively charged oxygen vacancy ($V_0^+$) is created, which acts as a donor of electrons. These positively charged oxygen vacancies can be moved in the direction of current by applying an electric field. Taking advantage of this ionic transport, the Hewlett-Packard team used a sandwich of thin conducting and nonconducting layers of $TiO_2$ to realize the memristor.

As shown in Figure 3(a), suppose we have two thin layers of $TiO_2$, one highly conducting layer with lots of oxygen vacancies ($V_0^+$) and the other layer undoped, which is highly resistive. Suppose that good Ohmic contacts are formed using platinum (Pt) electrodes on either side of this sandwich of $TiO_2$. The electronics barrier between the undoped $TiO_2$ and the metal looks broader, as shown. The situation remains the same, even when a negative potential is applied to electrode A, because the positively charged oxygen vacancies ($V_0^+$) are attracted towards electrode A and the length of the undoped region increases. Under these conditions, the electronic barrier at the undoped $TiO_2$ and the metal is still too wide and it will be difficult for the electrons to cross over the barrier. However, when a positive potential is applied to electrode A, as shown in Figure 3(b), the positively charged oxygen vacancies ($V_0^+$) are repelled and move into the undoped $TiO_2$. This ionic movement towards electrode B reduces the length of the undoped region. When more positively charged oxygen vacancies ($V_0^+$) reach the $TiO_2$/metal interface, the potential barrier for the electrons becomes very narrow, as shown in Figure 3, making tunnelling through the barrier a real possibility. This leads to a large current flow, making the device turn ON. In this case, the positively charged oxygen vacancies ($V_0^+$) are present across the length of the device. When the polarity of the applied voltage is reversed, the oxygen vacancies ($V_0^+$) can be pushed back into their original place on the doped side, restoring the broader electronic barrier at the $TiO_2$/metal interface. This forces the device to turn OFF due to an increase in the resistance of the device and reduced possibility for carrier

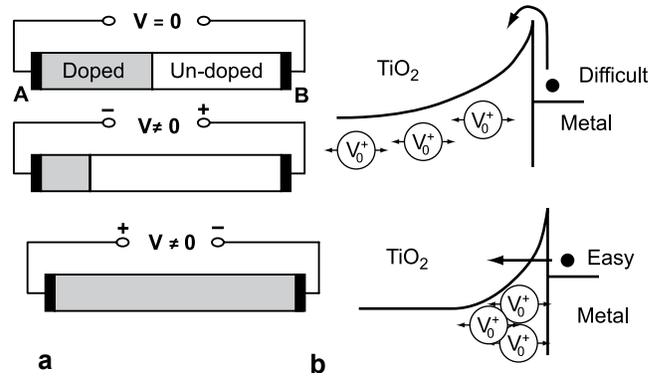

**Figure 3:** Conduction mechanism in a memristor (a) broader electronic barrier when a negative potential is applied to electrode A and (b) thin electronic barrier when a positive potential is applied to electrode A.

tunnelling [7].

The speciality of the memristor is not just that it can be turned OFF or ON but that it can actually remember the previous state. This is because when the applied bias is removed, the positively charged Ti ions (which are actually the oxygen deficient sites) do not move anymore, making the boundary between the doped and undoped layers of $TiO_2$ immobile. When you next apply a bias (negative or positive) to the device, it starts from where it was left. Unlike in the case of typical semiconductors, such as silicon in which only mobile carriers move, in the case of the memristor, both the ionic as well as the electron movement, into the undoped $TiO_2$ and out of undoped $TiO_2$, are responsible for the hysteresis in its current-voltage characteristics.

But one question remains in our mind. Where is the magnetic flux in this whole experiment carried out by the Hewlett-Packard team? In the beginning of our discussion on memristors, we said that the flux is supposed to be related to the charge through the memristance *M*. Fortunately, linking magnetic flux to the voltage is not the only way of satisfying the equation for memristance. There are other ways of satisfying the fourth fundamental equation, as long as the relationship between the integrals of the current and voltage is nonlinear [6]. For example, in the case of $TiO_2$ experiment, increasing or decreasing the length of the undoped region determines the resistance of the memristor. Therefore, here the state variable is the thickness of the stoichiometric or pure $TiO_2$ and it is controlled by the amount of charge passing through the device over a period of time [8]. Realization of this general nature of the fourth fundamental equation by the Hewlett-Packard team, after years of hard work, has made them recognize through their experiment that they have finally succeeded in building the memristor.

The future application of memristors in electronics is not very clear, since we do not yet know how to design circuits using memristors along with the silicon devices, although it is not difficult to integrate memristors on a silicon chip, since we now have matured technologies to accomplish this. However, since a memristor is a two terminal device, it is easier to address in a cross bar array. It appears, therefore, that the immediate application of memristors is in building nanoscale high density nonvolatile memories and field programmable gate arrays (FPGAs) [9-12]. Chua's original work also shows that using a memristor in an electronic circuit will reduce the transistor count by more than an order of magnitude. This may lead to higher component densities in a given chip area, helping us beat Moore's law.

Let me conclude by recollecting the prophetic statement made by Lean Chua in 1971: 'Although no physical memristor has yet been discovered in the form of a physical device without internal power supply, … a memristor device … could be invented, if not discovered accidentally. It is perhaps not unreasonable to suppose that such a device might already have been fabricated as a laboratory curiosity but was improperly identified!' [1]

Will this fundamental scientific discovery and its practical demonstration lead to a Nobel prize for the inventors? We do not yet know. Only time will reveal.

I look forward to hearing from you as always.

**M. Jagadesh Kumar**

Editor-in-Chief, IETE Technical Review,
Department of Electrical Engineering, IIT, Hauz Khas,
New Delhi - 110 016, India.
**E-mail:** mamidala@ieee.org